\documentclass{./styles/svproc}
\usepackage{url}

\usepackage{algorithmic} 
\usepackage{graphicx} 
\makeatletter
\renewcommand\inst[1]{\ifhmode\unskip\fi\ensuremath{^{#1}}}
\makeatother
\begin{document}
\mainmatter              
%

\title{RayPet: Unveiling Challenges and Solutions for Activity and Posture Recognition in Pets Using FMCW Mm-Wave Radar}
\titlerunning{RayPet}  
%
\author{Ehsan Sadeghi\inst{1} \and Abel van Raalte\inst{1}
 \and  Alessandro Chiumento\inst{1} \and Paul Havinga\inst{1} }
\authorrunning{Ehsan Sadeghi et al.} 
%
\tocauthor{Ehsan Sadeghi, Abel van Raalte, Alessandro Chiumento, and Paul Havinga}
\institute{\inst{1}Twente University, Enschede, The Netherlands.\\
\email{e.sadeghi@utwente.nl}
}
\maketitle 

\begin{abstract}

Recognizing animal activities (AAR) holds a crucial role in monitoring animals' health and well-being.
Additionally, a considerable audience is keen on monitoring their pets' well-being and health status. 
Insight into animals' habitual activities and patterns not only aids veterinarians in accurate diagnoses but also offers pet owners early alerts.
Traditional methods of tracking animal behavior involve wearable sensors like IMU sensors, collars, or cameras. Nevertheless, concerns, including privacy, robustness, and animal discomfort persist.
In this study, radar technology, a noninvasive remote sensing technology widely employed in human health monitoring, is explored for AAR. 
Radar enables fine motion analysis through Microdoppler spectrograms. 
Utilizing an off-the-shelf FMCW mm-wave radar, we gather data from five distinct activities and postures.
Merging radar technology with Machine Learning and Deep Learning algorithms helps distinguish diverse pet activities and postures.
Specific challenges in AAR, such as random movements, being uncontrollable, noise, and small animal size, make radar adoption for animal monitoring complex.
In this study, RayPet unveils different challenges and solutions regarding monitoring small animals.
To overcome the challenges, different signal processing steps are devised and implemented, tailored for animals. 
We use four types of classifiers and achieve an accuracy rate of 89\%. 
This progress marks an important step in using radar technology to observe and comprehend activities and postures in pets in particular and in animals in general, contributing to our knowledge of animal well-being and behavior analysis.

\end{abstract}

\section{Introduction}
Recently, recognizing different activities and postures using a variety of sensors gained a lot of attention. 
Initially, the healthcare sector became interested in human activity recognition (HAR) to monitor the elderly living alone and patients \cite{Intro1}. 
Later, researchers investigated animal activity recognition (AAR) to gain enough knowledge about the well-being of animals without manual vision-based scoring systems \cite{Intro2}.
Activity recognition (AR) is usually done using vision-based and sensor-based systems \cite{Paul}. 
Users often disregard vision-based systems due to privacy issues. 
Furthermore, vision-based systems are susceptible to variations in ambient light and environmental conditions. This vulnerability diminishes their robustness when considering their application in animal farming scenarios\cite{survey}.
Among sensor-based systems, wearable sensors are primarily utilized in both HAR and AAR.
However, studies show constantly wearing a sensor can be irritating and stressful, especially for animals \cite{survey}.

Considering all these challenges and concerns, radar, as a non-invasive sensing technology, has recently been utilized for different applications in human health monitoring, especially HAR. 
Feng Jin et al. (2019) used millimeter waves (mm-waves) frequency modulated continuous waves (FMCW) radar to build Doppler maps and spectrograms to detect different behaviors in patients using deep convolutional neural network (CNN) \cite{HAR1}.
Singh et al. (2019), in Radhar collected quite an extensive dataset from different human activities and used pre-processed point clouds (PCs) from mm-wave FMCW radar to differentiate between various types of activities \cite{HAR2}.
They used the voxelization technique prior to classification and achieved around 90\% accuracy in recognizing five different human activities.
Other researchers have utilized PCs collected from radar, Palipana et al. (2021) and Yu et al. (2022), for gesture and activity recognition, respectively.

Although radar has been widely utilized in different applications for human health monitoring, it has rarely been used in animal health monitoring.
Animals present additional challenges as compared to humans, there are many random movements and noises, which may make the processing steps more complex and decrease the system's accuracy\cite{survey}.
 With this in mind, a few researchers tried to apply radars to animals in different applications. 
Henry et al. (2018) used 24 GHz FMCW radar to detect sheep's position and motion \cite{AAR1}.
 Fioranelli et al. (2019) used FMCW radar working at 5.8 GHz to collect raw data, from micro-Doppler signatures, and perform lameness detection in ruminants, including dairy cows and sheep \cite{AAR2}.
Finally, Wang et al. (2020) investigated the vital sign detection of pets (dogs and cats) using ultra-wideband (UWB) radar. 
However, the data was collected while animals were at rest and not doing their routine activities and postures \cite{AAR3}.
 Even though there are limited studies on animals using FMCW radar, it sounds promising and has some potential for animal health monitoring applications \cite{survey}.

Among different animals, monitoring small animals presents heightened challenges \cite{survey}.
In RayPet we focus on small animals and the challenges associated with their monitoring.
To do so, we have chosen to concentrate on a readily accessible category of small animals: pets. This approach allows us to uncover and resolve the unique monitoring challenges associated with them.
Furthermore, it's prudent to monitor the well-being of pets, particularly because they often spend a long time alone at home.
Therefore, AAR serves as a valuable indicator of pets' welfare by quantifying the frequency and duration of their activities. 
While some wearable sensors like belts and collars are already available, wearing them can cause distress to small animals and has its own concerns and challenges, as discussed earlier.
As a result, we use a non-invasive mm-waves FMCW Radar as a remote sensing method for AAR.

In this article, our exploration delves into the application of AAR through the use of FMCW mm-wave radar. Our investigation centers on key research questions that have emerged in this domain. These questions encompass discerning the distinctions between HAR and AAR, unveiling the complexities linked to recognizing small animal activities and postures, and formulating strategies for designing and executing signal preprocessing steps aimed at resolving these differences and challenges.
As far as our current understanding goes, we are pioneering the application of mm-wave FMCW radar for AAR.
Our goal is to uncover challenges associated with employing FMCW mm-wave radar for small animal activity and posture recognition. Subsequently, by addressing both animal-related issues and challenges pertaining to point cloud sparsity, quantity, and quality, we formulate viable solutions.
Therefore, in RayPet, by focusing on pets, we gather the raw data points, execute necessary implementations, and devise pre-processing methods aimed at enhancing the acquired dataset to optimize performance for our classification algorithms. 
In this study, we investigate different trade-offs in pre-processing methods and tailor processing techniques like Noise Removal to the requirements for AAR and its challenges.  
Finally, we compare our results with Radhar \cite{HAR2}, which is designed for HAR, to indicate and emphasize the effectiveness of the approach for animals.

The rest of the paper is organized as follows.
In Section II, we give a brief overview of mm-wave FMCW radar and propose our designed system model. 
Section III describes the signal pre-processing steps, including Noise Removal, Data Aggregation, and Voxelization.
Section IV explains the designed setup for data collection, signal pre-processing parameters and trade-offs, and models and algorithms used for classification.
Later on, in Section V, the output of our algorithms is evaluated and discussed, considering different signal pre-processing blocks.
Finally, Section VI concludes the paper and reviews some opportunities to improve the designed system.

\section{System Model}

\subsection{Overview of FMCW Radar}

\begin{figure}[htbp]
\centerline{\includegraphics[width=95mm,scale=1]{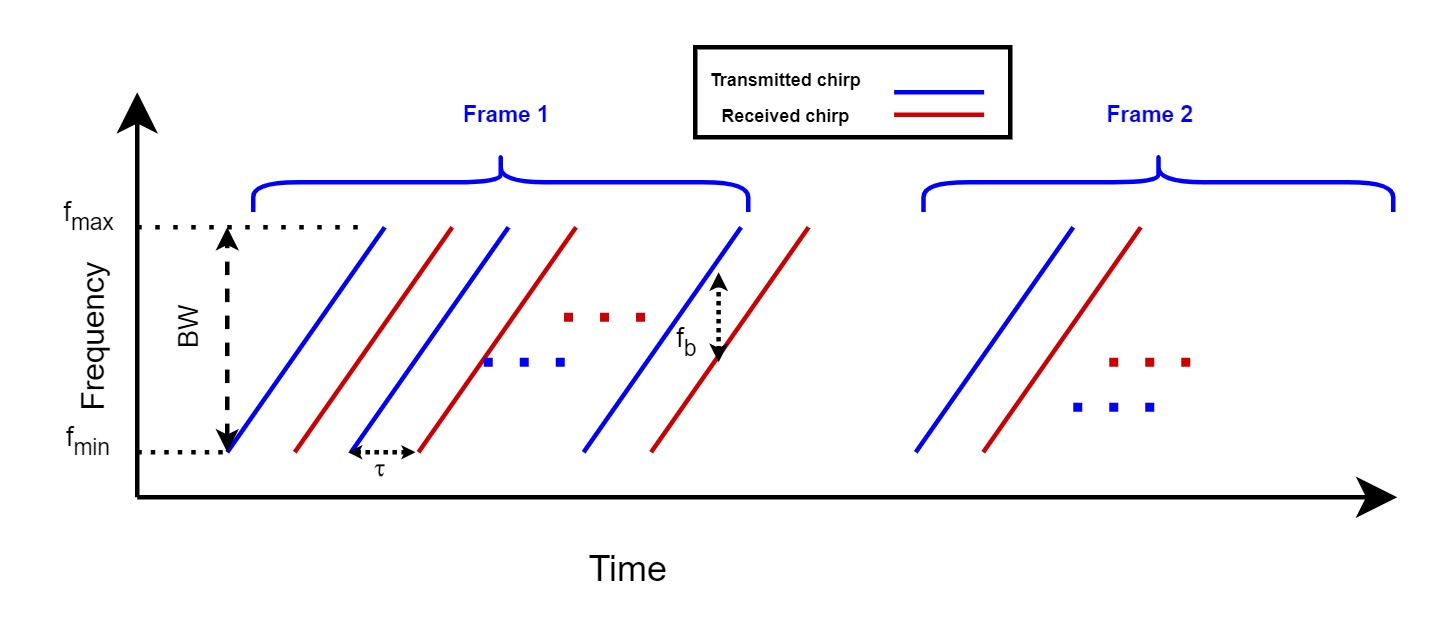}}
\vspace{-1.5em}
\caption{Frequency changes in transmitted and received chirps in FMCW radar. }
\label{FMCWradar}
\end{figure}

Echolocating mammals inspire the concept of radars. Echolocating mammals send sound waves away and listen to the reflection to localize different objects and detect their distances. 
Radars, similarly, send electromagnetic waves away and listen to reflected signals of the object/ subject. 
In FMCW radar, the transmitted sinusoidal signal's frequency linearly changes over time, and it contains multiple chirps in each frame, as depicted in Figure \ref{FMCWradar}. 
The figure shows that the frequency is swept between $f_{min}$ to $f_{max}$. 
We call $f_{max}-f_{min}= BW$, the sweeping bandwidth (BW). 
The received chirps are the delayed version of the transmitted chirps.
$\tau$ also indicates the delay between the transmitted and received chirps.
$\tau= \frac{2\times d}{C}$, in which d is the distance of the object from radar ($ d= \frac{\tau}{2 C}$) and C is the light speed.

Beat frequency or $f_{b}$ is the frequency difference between the transmitted and received chirps. The output of radar is called intermediate frequency signal (IF signal) with $f_{b}$ as its frequency, which is generated after combining the transmitted and received signals. Later, this signal will be sampled  and processed to achieve point clouds of the object.
To achieve x, y, z coordinates, different processing steps like range FFT, Velocity FFT, constant false alarm rate (CFAR), and angle FFT should be implemented \cite{TI}.
The radar capability in distinguishing between two different objects are shown by range resolution ($\Delta R$), which can be calculated as follows:

\begin{equation}
\Delta R = \frac{C}{2 BW}.
\end{equation}

It is worth mentioning that we are using Texas Instruments (TI) FMCW mm-Wave radar. Mm-waves signals work in the range of 30 to 300 (GHz), and it has various advantages. For instance, it has a higher attenuation than the other frequency ranges lower than 30 GHz, which helps the user experience isolated private sensing.
Furthermore, its shorter wavelength facilitates the creation of compact antenna patches, allowing for monitoring intricate and detailed activities within a confined space.

\begin{figure*}[htbp]
\centerline{\includegraphics[width=14cm,scale=2]{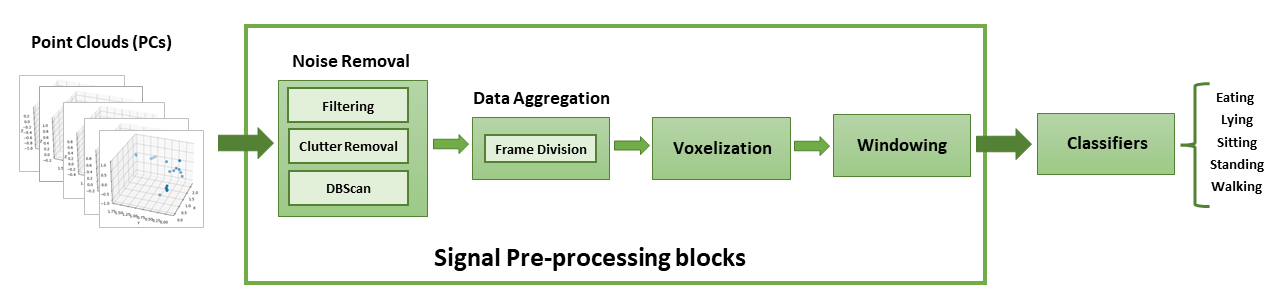}}
\vspace{-0.5em}
\caption{System overview and pre-processing blocks. }
\label{sys}
\end{figure*}

\subsection{Overview of System Design}
In this article, RayPet aims at processing and using FMCW mm-wave radar's point clouds (PCs) for activity and posture recognition in dogs. 
The architecture of our system contains two main blocks, namely signal pre-processing and classification, and multiple sub-blocks, shown in Figure \ref{sys}.
Point cloud representation, which is collected by radar and is the input of our system, consists of different reflected points from the object or scene area. The point cloud representation contains information like (x, y, z) coordinates, velocity, and intensity of the points.
Although radar IF signal goes through multiple steps before preparing these PCs, there are still some issues and challenges in the quality of the collected PCs.
First, the limited number of Pcs collected by radar and its sparsity can decrease the quality of the point cloud representation.
Second, PCs are vulnerable to different factors, like multi-path propagation, scene area, environmental effects, interference, etc., which can significantly degrade the quality and accuracy of PCs.
Third, the quality of PCs in animal activity recognition (AAR) is lower as compared to human activity recognition (HAR). 
That is because, in AAR, we have many random movements done by animals that can be irrelevant to the performing activity, like wagging the tail in dogs.

All these challenges may ultimately lead to a decrease in the system's overall performance. 
As a result, we designed different signal pre-processing sub-blocks in our system to cope with these challenges.
It is worth mentioning that we tailored these pre-processing steps to suit the unique needs and characteristics of the animals.
In Figure \ref{sys}, the overview of the proposed and designed system to cope with these challenges is depicted.
As shown in Figure \ref{sys}, the collected PCs first go through the noise removal step, in which filtering, clutter removal algorithms, and Density-based spatial clustering of applications with noise (DBScan) are applied.
Subsequently, it goes through the data aggregation step. Then, the data will be voxelized to prepare for input into the classifiers. Later on, windowing techniques are applied to the voxelized point cloud. Finally, the data is fed into the classifiers to distinguish among activities and postures.
In the following, we elaborate on these signal-processing steps and their use cases.

\section{Signal Pre-processing}

In this section, we will examine various pre-processing steps designed to enhance the quality of PCs while considering animal-specific features. As was mentioned earlier, the quality of the PCs collected from an animal is less than HAR applications due to various reasons like sparsity of the PCs, random body movements in animals, etc. 
As a result, we designed and implemented multiple signal pre-processing blocks to improve the point cloud representation quality to some extent.

\subsection{Noise Removal}
This block primarily aims to eliminate noisy points, decrease sparsity, eliminate outliers, and remove scattered points from static objects. To do so, we designed different signal pre-processing sub-blocks, which will be discussed in the following.

\paragraph{Filtering}
To reduce the number of noisy points appearing in the point cloud representation, we first collect the point cloud representation of the scene area without the object. Doing so makes it easier to apply a filter and detect noisy points in the point cloud representation of each specific animal activity.

\paragraph{Static Clutter Removal}

Static clutter removal is a signal processing method to eliminate unwanted points (noise) and clutters from the dataset, especially when the noise remains stationary over time. Ultimately, the static clutter removal algorithm identifies and filters out the noisy points without any remarkable changes or motion over successive chirps or frames. 
The critical point in static clutter removal algorithm implementation is to evaluate the temporal consistency of PCs and distinguish between static clutter and the dynamic object. Note that even if the posture or behavior we are looking into is almost static, we notice changes in living creatures over time. Assuming noise (from the static objects) remains stationary while relevant objects exhibit motions, static clutter removal can clean the data set and eliminate the static noisy points from the PC. By doing so, we have fewer points resulting from sensor inaccuracy or background interference, etc., and we prepare a better dataset for our classifiers.

\paragraph{DBScan}

Density-Based Spatial Clustering of Applications with Noise (DBScan) is a clustering algorithm that groups points based on their density rather than considering predefined cluster shape.
DBScan identifies core points with sufficient neighboring points within a specified distance (eps) and expands clusters by linking neighboring core and border points. 
In the end, it forms different clusters with various numbers of points. 
Points not belonging to any cluster are marked as noise or outliers.
In some studies, researchers detect the main cluster and remove outliers together with all other clusters \cite{HAR3}. 
However, due to the small size of animals, there is a high chance we remove clusters related to legs or head, which can involve unique patterns of PCs for each specific activity or posture.
Considering this, we only use DBScan to remove outliers and noisy points to prevent losing information.
Outliers can be resulted from background noises or random irrelevant movements in the dog, for example, tail wagging which has nothing to do with the ongoing activity or posture.
Different parameters should be set for DBScan, like Epsilon (eps) and minimum points in each cluster. 
Note that we applied DBScan on each frame collected PCs and performed parameter fine-tuning using a grid search to set the required parameters.

\subsection{Data Aggregation}

The number of collected points in each frame is usually limited by radar and constant false alarm rate detection (CFAR) performance. 
The low number of points can result in poor performance and low accuracy in classifiers.
With this in mind, we implement data aggregation as an algorithm to increase the number of collected points in each frame and end up with a better, cleaner point cloud representation.
In data aggregation, we aggregate the collected points from multiple frames with each other and form new frames.
By doing so, we have more points in each frame and a higher chance of getting a better result.
Although data aggregation can improve the quality of PCs collected in each frame significantly, it reduces datasets (as we have a constant number of voxels which will be explained later).
Therefore, there is always a trade-off that should be considered; decreasing the dataset size may lead to underfitting in the classifiers. 
Furthermore, taking into account the frame duration, there exists a constraint on the maximum number of frames we can aggregate. If the total duration of the aggregated frames surpasses a predefined limit (determined by the movement of objects and their speed), those aggregated frames may not pertain to the same thing, rendering the entire approach ineffective.

\subsection{Voxelization}

Data collected in each frame contains several point clouds with different attributed values.
Consequently, the features utilized for classifiers exhibit varying sizes and formats.
As a solution, we employ the voxelization method on each frame to ensure a consistent feature size, making it suitable for integration with diverse classifier models.
Voxelization is a data processing method that converts three-dimensional data into a structured grid of voxels (cubes).
We consider a constant number of voxels $m \times n \times p$. 

We should note that each data point obtained from the radar includes essential details such as its spatial position in three-dimensional space denoted by (x, y, z) coordinates, its velocity, which describes the speed, and an intensity value that reflects the strength of the radar signal returned by the point.
The value of the voxels can be determined based on the coordinates themselves, velocity, or intensity. 
We put all these options to the test, and it turned out the coordinates were the best option in terms of performance and accuracy.
Therefore, in our voxelization method, each cube or voxel is assigned a number if a point exists.
In non-voxelized data, every point within a 3D grid is assigned individual x, y, and z coordinates. However, cubic elements are employed in the voxelized representation of the same data, each capable of containing none, one, or multiple points.

\subsection{Windowing}

Finally, windowing is applied to the voxelized representation of data before feeding them into classifiers. 
Windowing helps us having a smoother transition over the data set, building features for classifiers and maintaining the temporal context of the original data. 
Considering all these, we apply windowing to preserve the temporal changes during different activities and postures.
We should note that there is a limit to the window size considering each posture or activity duration.
Also, there is always a trade-off between windowing parameters and data size.
We set $W$ and $SW$, as window size and sliding window factor.
Sliding windows or overlapping windows can improve the temporal resolution of the analysis.
As the size of each voxel is  $m \times n \times p$, the feature size fed to the classifier has a shape of  $W \times m \times n \times p$.

In the following, we explain our experiments and parameters in more detail and put our proposed approach for AAR to the test.

\section{Experiment}
\subsection{Radar Configuration and data acquisition}

In this experiment, we used a commodity IWR1443  Evaluation Module (EVM) which is a mm-wave FMCW radar \cite{TIIWR1443}.
IWR1443 EVM operates on 76 to 81 GHz frequency band (up to 4 GHz bandwidth) and has three transmitters (TX) antennas and four receivers (RX) antennas, which enables calculating both azimuth and elevation angles which is useful for generating the PCs.
As we discussed earlier, many parameters should be configured regarding the application. In Table \ref{tab1}, the summary of the radar configuration used in this study is listed. By implementing this setup, we can attain an approximate range resolution of 0.05 meters.

\begin{table}
\caption{Radar configuration}
\begin{center}
\begin{tabular}{c@{\quad}lr}
\hline
Parameters&Values\\
\hline
No. of Samples &  240 \\
No. of chirps &  16 \\
Starting frequency & 79.21 GHz   \\
Frame duration &  33.33 ms \\
Bandwidth (BW) &  2439.8 MHz\\
Pulse repetition interval (PRI) &  64.140 $\mu$ \\
\hline
\end{tabular}
\label{tab1}
\end{center}
\end{table}



Raw data should go through multiple steps to collect three-dimensional PCs using IWR1443 FMCW mm-wave radar, including range map formation, doppler map formation, angle estimation, and CFAR.
To implement this, we used Robot Operating System (ROS), an open-source framework that provides a collection of software libraries and tools.
It facilitated essential steps such as generating point cloud data and visualizing these PCs in real-time.

\subsection{Experimental setup and data collection}

\begin{figure}[htbp]
\centerline{\includegraphics[width=65mm,scale=0.8]{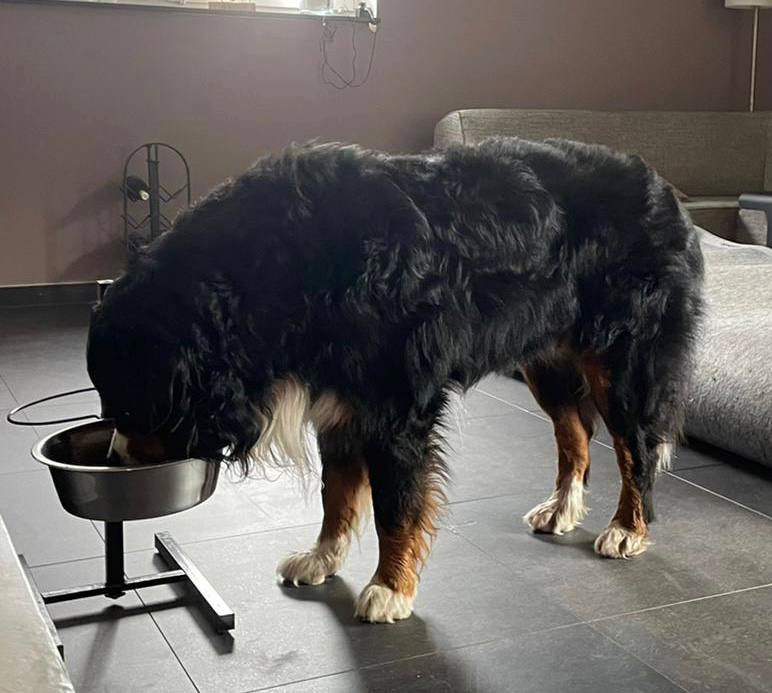}}
\vspace{-0.5em}
\caption{The experimental setup: The dog doing the eating activity in the scene area. }
\label{dog}
\end{figure}

In this experiment, we collected data from a dog. 
The dog's physical characteristics include a weight of approximately 45kg, a height of 0.75 meters, and a length of 1.10 meters. 
Given these attributes, the radar's height was set at 0.50 meters.
Five different activities and postures, namely eating, lying, sitting, standing, and walking, were performed at a distance of around 1.30 meters from the radar.
Compared to HAR, AAR presents additional challenges in maintaining animals in specific positions for data collection.
Furthermore, allowing the animals to continue their regular daily activities throughout the recording is crucial.
To address this challenge, we adapted our data collection approach by recording data in shorter intervals of 10 seconds (5 seconds for walking as it takes shorter for the dog to pass radar beam width). Although this resulted in a longer data collection procedure, it allowed us to accommodate the animals' natural movements and behaviors more effectively. 
Ultimately, we managed to record over 2200 seconds for all activities. For instance, in Figure \ref{dog}, you can see the dog while doing the eating activity.

\subsection{Pre-processing}

As previously outlined in the pre-processing steps, the gathered data must undergo various signal pre-processing stages to be prepared for classification.
Initially, noise elimination was executed by implementing noise removal methods, filtering, clutter removal, and DBScan.
When configuring DBScan parameters, we undertook assessments using different values that aligned with point density and activities. Ultimately, through empirical analysis, we set the parameters eps and the minimum number of points to (0.5, 2). 
By doing so, we eliminated a significant portion of noise in our dataset.

Following this, the refined dataset underwent data aggregation via data aggregation. 
We experimented with aggregating diverse frame quantities (2, 3, 4, 5). 
It is crucial to note that the increased number of concatenated frames may potentially result in losing temporal changes.
Furthermore, aggregating a larger number of frames into a single frame could potentially yield smaller datasets, potentially leading to underfitting within the classifiers.
In light of these considerations, we opted to merge only two frames during the data aggregation phase. This decision was made because, when merging more than two frames, we observed a noticeable decrease in the overall system accuracy, which could be indicative of underfitting.
The new modified data set should now go through voxelization.
We explored varying voxel quantities in each dimension, considering the animal size and the scope of movement.
However, the dimension used by other studies on HAR ($10\times32\times32)$ turned out to be best for our application.

Finally, for windowing we investigated different values for $W$ and $SW$ such as (W=20, SW=4), (W=25, SW=5),(W=30, SW=10), and (W=45, SW=10).
While windowing undoubtedly enhances feature quality for classifiers, selecting appropriate windows and sliding window sizes hinges on various factors.
For instance, excessively enlarging the window size could result in a reduced number of training examples, potentially causing the loss of spatial and temporal changes within the dataset.
Within the set of activities and postures we are focusing on, there are two activities and three postures. 
Activities are walking, dynamic motion, and eating, which involves intricate movements around the neck and mouth. 
Preserving the temporal dependency holds significant importance in accurately capturing these two activities.
Therefore, we select windowing and sliding parameters in a way to enhance the precision of our estimations for these actions and elevate the overall accuracy.
It is imperative to note that opting for a larger window size may eventually result in a reduced number of available training examples, potentially leading to underfitting. We visually illustrate this delicate trade-off in a figure that will be properly explored in the upcoming discussion section.
Having thoroughly evaluated various window and sliding window sizes, we determined that setting W=30 and SW=10 yielded enhanced performance.
Consequently, the classifier is presented with features structured and shaped in the format of $30 \times 10 \times 32 \times 32$.

\subsection{Classifiers and Model Parameters}

Prior to classification and activity recognition, we split our dataset into two portions: 
30\% for testing and 70\% for training. 
We employed four distinct classifiers on our dataset to implement AAR. 
These classifiers include Support Vector Machine (SVM) with Principal Component Analysis (PCA), Multi-layer Perception (MLP), Bidirectional Long Short-term Memory (Bi-LSTM), and Time Distributed Convolutional Neural Network with Bi-LSTM (TD-CNN+Bi-LSTM). 
In the following, we delve into the details of these classifiers, the respective models, and the chosen classification parameters.

\paragraph{SVM+PCA}

We begin with the initial classifier, namely the Support Vector Machine (SVM).
The voxelized representation, initially structured as $30 \times 10 \times 32 \times 32$, is flattened, resulting in a data dimension of 307200.
Given the computational inefficiency associated with such extensive features, we employed Principal Component Analysis (PCA) to reduce the data's dimensionality while preserving its variability.
This endeavor led to a reduction from 307200 dimensions to 3000 dimensions.
The radial basis function (RBF) served as the utilized Kernel function.
The implementation of SVM, coupled with PCA, was executed through Keras. Subsequently, for the optimization of SVM hyperparameters (C and gamma), we utilized GridSearchVC.
Adam optimizer with a learning rate of 0.001 was used to update the network's parameters during training, aiming to reach the optimal set of weights and biases.

\paragraph{MLP}

The Multi-Layer Perceptron (MLP) represents a form of artificial neural network (ANN) characterized by multiple tiers of interconnected nodes or neurons.
In the context of this investigation, our focus lies on four fully connected layers.
Just as with the SVM, the voxelized representation of the data undergoes flattening, thus giving rise to an input dimension of 307200.
The realization of this architecture involves the employment of both Keras and SKlearn. Ultimately, after undergoing 40 training epochs, the model exhibiting the lowest validation loss is selected.
\paragraph{Bi-LSTM}

Comprising dual LSTM layers, the Bidirectional Long Short-term Memory (Bi-LSTM) takes into account both past and future context when processing sequences.
In contrast, a conventional LSTM handles sequences in a unidirectional manner.
Bi-LSTM allows the network to capture past and future information at each time step, potentially improving its ability to model complex dependencies in the data.
Considering this, Bi-LSTM can be a good approach for capturing temporal dependencies in datasets representing an activity or posture.
To achieve this, we flatten the spatial dimension within the data (voxel representation of $10 \times 32 \times 32 = 1024$) and capture the temporal patterns (window/ time dimension). As a result, the input shape for the Bi-LSTM model, with 64 units in size and 64 hidden units, takes on the shape of ($30 \times 1024$).
Likewise, we employed Keras and SKlearn to construct the classifier and employed the Adam optimizer with a learning rate of 0.001 for training the classifier. Following 40 training epochs, the model demonstrating the lowest validation loss was identified and retained.

\paragraph{TD-CNN+Bi-LSTM}

Integrating a Time Distributed Convolutional Neural Network (TD-CNN) with a Bi-LSTM constitutes a robust technique for processing sequential data that incorporates both spatial and temporal attributes. 
As a result, it is an adequate solution for categorizing diverse activities and postures where both temporal and spatial variations play a pivotal role.
In this instance, the TD-CNN + Bi-LSTM model undergoes training utilizing the voxelized representation, encompassing both temporal and spatial dimensions. 
Similarly, we employ Sklearn and Keras to implement the classifier and employ the Adam optimizer with a learning rate of 0.001 for training purposes. The model exhibiting the least loss is selected after completing 40 training epochs.

\section{Discussion}

We fed voxelized representations of five distinct activities, as well as the flattened version, into the four distinct classifiers. 
Prior to this, as previously mentioned, we divided the dataset into separate training and test sets to prevent any data leakage.
The overall accuracy of SVM (PCA=3000), MLP, Bi-LSTM, and TD-CNN+Bi LSTM were 41\%, 78\%, 79\%, and 89\%, respectively.
To indicate the significance of our work, we employed the approach outlined by Radhar \cite{HAR2} as one of the pioneering and highly noted studies focusing solely on windowing and voxelization for HAR. 
We then compared this method with our own in Figure 4.
Evidently, our approach demonstrates a marked superiority over the alternative method.
This performance gap underscores our consideration of diverse challenges and physiological nuances in animals. 
In response, we devised a tailored signal processing methodology, strategically incorporating specific steps to address these factors comprehensively.

\begin{figure}[htbp]
\centerline{\includegraphics[width=85mm,scale=1]{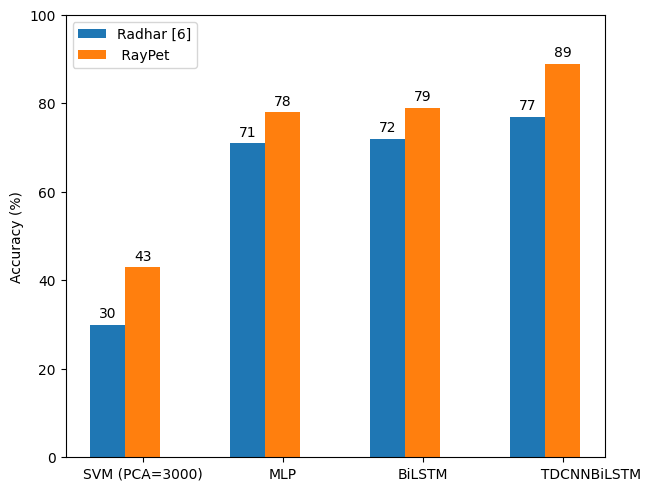}}
\vspace{-1em}
\caption{Highlighting the efficacy of RayPet through comparison with Radhar. }
\label{comparison}
\end{figure}

\begin{figure}[h]
\centerline{\includegraphics[width=85mm,scale=1]{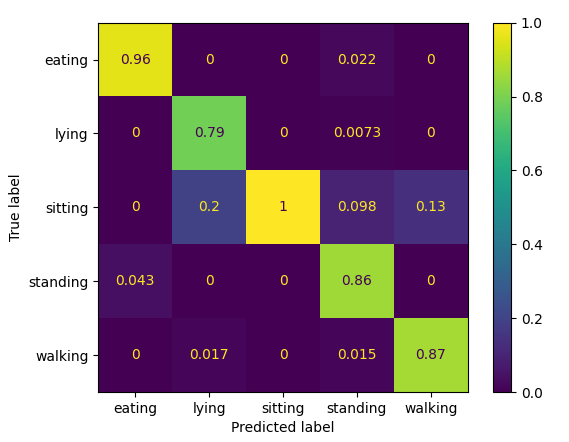}}
\vspace{-1em}
\caption{Confusion Matrix }
\label{ConfusionMatris}
\end{figure}

Among the classifiers we put to the test, SVM exhibited the poorest performance. 
That is because we flattened the voxelized representation before feeding it into the classifier.
By doing so, SVM is not utilizing any temporal or spatial dependencies within the voxelized dataset representation.
While the rest of the deep learning classifiers exhibit superior performance compared to SVM, it appears that MLP, in comparison to the others, displays a relatively weaker performance. 
It is essential to highlight that even though MLP is trained using the voxelized representation, no prior presumptions regarding the temporal and spatial dimensions within the data were incorporated into the MLP model.
Bi-LSTM performs slightly more effectively because it retains the temporal dependencies within the dataset and sequences. 
Processing sequences in dual directions can capture both temporal dependencies and complex patterns present within sequential data.
Among our classifiers, TD-CNN + Bi-LSTM stands out as notably superior to the rest.
The combination of TD-CNN and Bi-LSTM allows us to capture both spatial and temporal dependencies and patterns within the dataset.
This is essential in applications like activity or posture recognition in which the PCs (spatial information) or pattern evolves over time.
In Figure 5, we created the confusion matrix for our method using TD-CNN + Bi-LSTM, which had an overall accuracy of 85\%.
The activities and postures of lying down and sitting were frequently misclassified. This misclassification can be attributed to the inherent similarity between these activities and postures and the lack of adequate point cloud density for differentiation.

\begin{figure}[h]
\centerline{\includegraphics[width=85mm,scale=1]{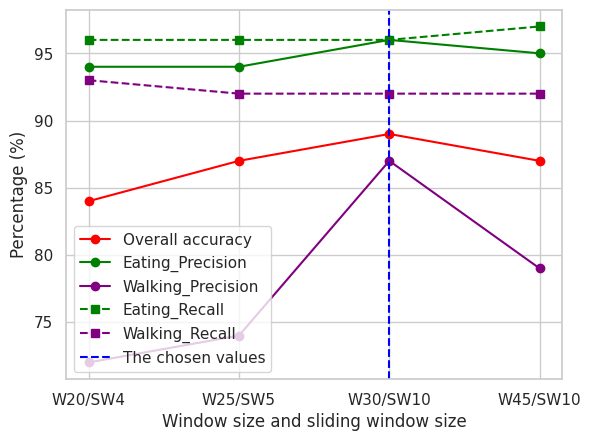}}
\vspace{-1em}
\caption{Navigating the Trade-off in Windowing and Sliding Window Sizes}
\label{Tradeoff}
\end{figure}

After testing the classifiers, the trade-off in choosing the right window and sliding window size is worth investigating.
As discussed previously, opting for extended window sizes plays a pivotal role in preserving the intricate temporal dependencies inherent in activities such as walking and eating.
Figure 6 shows that augmenting the window size, up to a limit of 30, contributes significantly to the recall and accuracy of activities like walking and eating.
What is more, our system achieves its peak overall accuracy at a window size of 30.
However, exceeding this threshold by increasing the window size further results in a decline in these metrics, possibly indicating the underfitting—as we discussed earlier.
Taking into account all these factors, a window size of 30 emerges as the optimal choice for our system's performance.

The study encountered several challenges and limitations that warrant discussion. Firstly, voxelization resulted in expanded dataset dimensions, necessitating increased computational resources for Deep Learning algorithm training. 
Secondly, acquiring training data for classifiers posed considerable difficulty in the case of animals, given their tendency to refuse maintaining consistent postures or activities.
Thirdly, even though we reduced animal-related noise side-effects to a great extent, there are still some animal-specific traits and random movements which cannot be indicative of any specific activity. A prime example is observed in dogs when they wag their tails.
Lastly, a significant challenge arises from radar and hardware limitations, which impose constraints on the quantity and quality of points that can be gathered via the radar system. 
Furthermore, it is important to note that the density of the point clouds for AAR doesnot align with that of HAR.
At times, the sole distinguishing factor between various postures or activities lies in the stance of the legs. 
Nevertheless, collecting a satisfactory quantity of points reflecting off the leg area proves to be considerably challenging due to limitations stemming from both the animal size and hardware limitations.

In conclusion, two potential strategies could be implemented to enhance the system's performance in animal activity recognition.
The first approach involves leveraging micro-Doppler signatures to capture intricate movements. This technique could significantly enhance the system's ability to detect fine motions and behaviors.
The second approach entails deploying multiple radars positioned at different locations. This strategy offers spatial diversity, augmenting both the quantity and quality of point cloud data. This, in turn, could lead to more accurate and robust activity recognition results.
Moving forward, RayPet is committed to exploring these avenues in forthcoming studies.

\section{Conclusion}

In animal sensing technologies, the prevailing methodologies often necessitate animals to wear sensors continually, encompassing wearable options such as IMUs or collars. 
However, RayPet takes a distinctive approach by leveraging FMCW radars—a noninvasive sensing technology—for various applications like activity and posture recognition. 
While radar technology holds immense potential, its application to animals requires careful consideration due to the radar's sensitivity, the smaller size of animals compared to humans, and the prevalence of random movements in animals.
Throughout our study, we tried to answer the mentioned research questions comprehensivly.  
In RayPet, we Unveiled and discovered different challenges involved in applying FMCW mm-wave radar to animals. 
As a solution, we developed and implemented distinct signal processing methods, encompassing noise removal algorithms, data aggregation, windowing, and voxelization. 
These procedures were crucial for effectively preprocessing and refining datasets before feeding into the classifiers.
This method proved particularly effective in accommodating even small animals like dogs.
To underscore the significance of our proposed system model and signal processing methods, we rigorously tested our system and compared it with an outstanding method specifically designed for HAR.
Our evaluation encompassed five distinct activities, increasing of up to 12\% in overall accuracy.
Looking ahead, our radar-based system model holds more promising potential with the incorporation of multiple radars simultaneously.

%
%


\begin{thebibliography}{6}
%

\bibitem{Intro1} Orces, Carlos H. "Prevalence and determinants of falls among older adults in Ecuador: an analysis of the SABE I survey." Current gerontology and geriatrics research 2013 (2013).



\bibitem{Intro2} Mullins, Israel L., Carissa M. Truman, Magnus R. Campler, Jeffrey M. Bewley, and Joao HC Costa. "Validation of a commercial automated body condition scoring system on a commercial dairy farm." Animals 9, no. 6 (2019): 287.


\bibitem{Paul} Bosch, Stephan, Filipe Serra Bragança, Mihai Marin-Perianu, Raluca Marin-Perianu, Berend Jan Van der Zwaag, John Voskamp, Willem Back, René Van Weeren, and Paul Havinga. "Equimoves: A wireless networked inertial measurement system for objective examination of horse gait." Sensors 18, no. 3 (2018): 850.

\bibitem{survey} Sadeghi, Ehsan, Claudie Kappers, Alessandro Chiumento, Marjolein Derks, and Paul Havinga. "Improving piglets health and well-being: a review of piglets health indicators and related sensing technologies." Smart Agricultural Technology (2023): 100246.







\bibitem{HAR1} Jin, Feng, et al. "Multiple patients behavior detection in real-time using mmWave radar and deep CNNs." 2019 IEEE Radar Conference (RadarConf). IEEE, 2019.


\bibitem{HAR2} Singh, Akash Deep, Sandeep Singh Sandha, Luis Garcia, and Mani Srivastava. "Radhar: Human activity recognition from point clouds generated through a millimeter-wave radar." In Proceedings of the 3rd ACM Workshop on Millimeter-wave Networks and Sensing Systems, pp. 51-56. 2019.

\bibitem{HAR3} Palipana, S., Salami, D., Leiva, L. A., and Sigg, S. (2021). Pantomime: Mid-air gesture recognition with sparse millimeter-wave radar point clouds. Proceedings of the ACM on Interactive, Mobile, Wearable and Ubiquitous Technologies, 5(1), 1-27.


\bibitem{HAR4}
Yu, Chengxi, Zhezhuang Xu, Kun Yan, Ying-Ren Chien, Shih-Hau Fang, and Hsiao-Chun Wu. "Noninvasive human activity recognition using millimeter-wave radar." IEEE Systems Journal 16, no. 2 (2022): 3036-3047.



\bibitem{AAR1} Henry, Dominique, Hervé Aubert, Edmond Ricard, Dominique Hazard, and Mathieu Lihoreau. "Automated monitoring of livestock behavior using frequency-modulated continuous-wave radars." Progress In Electromagnetics Research M 69 (2018): 151-160.


\bibitem{AAR2} Fioranelli, Francesco, Haobo Li, Julien Le Kernec, Valentina Busin, Nicholas Jonsson, George King, Martin Tomlinson, and Lorenzo Viora. "Radar-based evaluation of lameness detection in ruminants: preliminary results." In 2019 IEEE MTT-S International Microwave Biomedical Conference (IMBioC), vol. 1, pp. 1-4. IEEE, 2019.



\bibitem{AAR3} Wang, Pengfei, Yangyang Ma, Fulai Liang, Yang Zhang, Xiao Yu, Zhao Li, Qiang An, Hao Lv, and Jianqi Wang. "Non-contact vital signs monitoring of dog and cat using a UWB radar." Animals 10, no. 2 (2020): 205.




\bibitem{TI} C. Iovescu and S. Rao, “The fundamentals of millimeter wave sensors,” Texas Instruments, 2017.


\bibitem{TIIWR1443} IWR1443, , “IWR1443 Single-Chip 76 to 81-GHz mmWave Sensor
Datasheet,” Texas Instrum., Dallas, TX, USA. Accessed: Aug. 03, 2023.
[Online]. Available: https://www.ti.com/lit/gpn/iwr1443.



\end{thebibliography}
\end{document}